\documentclass[aps,twocolumn,pra,showpacs,amsmath,amssymb,showkeys,10pt]{revtex4-2}
\usepackage{graphicx}
\usepackage{epstopdf}
\pdfoutput=1
\usepackage{braket}
\usepackage{hyperref}
\usepackage{longtable}

\begin{document}

	\title{Comparing the effects of nuclear and electron spins on the formation of neutral hydrogen molecule}
	
	\author{Miao Hui-hui}
	\email[Email address: ]{miaohuihui3@gmail.com}
	\affiliation{Faculty of Computational Mathematics and Cybernetics, Lomonosov Moscow State University, Vorobyovy Gory, Moscow, 119991, Russia}

	\author{Ozhigov Yuri Igorevich}
	\email[Email address: ]{ozhigov@cs.msu.ru}
	\affiliation{Faculty of Computational Mathematics and Cybernetics, Lomonosov Moscow State University, Vorobyovy Gory, Moscow, 119991, Russia\\K. A. Valiev Institute of physics and technology, Russian Academy of Sciences, Nakhimovsky Prospekt 36, Moscow, 117218, Russia}

	\date{\today}

	\begin{abstract}
	We introduce the association-dissociation model of neutral hydrogen molecule, which is a finite-dimensional cavity quantum electrodynamics model of chemistry with two two-level artificial atoms on quantum dots placed in optical cavities, based on the Tavis--Cummings--Hubbard model. The motion of the nuclei can be represented in quantum form. Electron spin transition and spin spin interaction between electron and nucleus are both considered. Consideration is also given to the effects of nuclear and electron spins on the formation of neutral hydrogen molecule.
	\end{abstract}


	\maketitle

	\section{Introduction}
	\label{sec:Intro}

	The modelling of hydrogen chemical processes attracts increasing interest and becomes one of the primary tasks in recent years, including chemical reactions involving cation $\mathrm{H}_2^+$ \cite{Zhu2020, Afanasyev2021} and neutral hydrogen molecule $\mathrm{H}_2$ \cite{Miao2023}. Quantum chemistry is usually understood as a technique for calculating the numerical characteristics of stationary atoms or molecules: binding energies, spectra, etc. This paper is devoted to a different direction: the dynamics of chemical reactions and the influence of the electromagnetic field and the thermal properties of the environment on them. The task of describing dynamic reaction scenarios is very demanding in terms of computational resources, and therefore incompatible with the exact calculation of the characteristics of stationary structures. We assume that the exact values of the binding energies, electron tunnelling and their interaction with the field are test parameters that can be determined not only by standard computational methods (Hartree-Fock, Monte Carlo and density functional), but also selected from observing the outcomes of dynamic association-dissociation scenarios, the mechanisms of which we are building. This paper provides a method for extending the cavity quantum electrodynamics (QED) model to complex chemical and even biological models by studying the association-dissociation reaction of hydrogen molecule. This is significant because this model can be modified in the future for use with more intricate chemical and biological models, which necessitate an understanding of hydrogen chemical processes. In this paper, the association-dissociation model of neutral hydrogen molecule is introduced in detail, and the effects of nuclear and electron spins on the formation of neutral hydrogen molecule is compared.
	
	The most commonly used cavity QED models are the Jaynes-Cummings model (JCM) \cite{Jaynes1963} and the Tavis-Cummings model (TCM) \cite{Tavis1968}, describing the dynamics of one or a group of two-level atoms in an optical cavity, which are the fundamental models for strong coupling (SC). JCM and TCM have been generalized to several cavities coupled by an optical fiber --- the Jaynes-Cummings-Hubbard model (JCHM) and Tavis-Cummings-Hubbard model (TCHM) \cite{Angelakis2007}. The value of these models and their modifications is that it allows us to describe a very complex interaction of light and matter in the framework of finite-dimensional QED models. Recently, a lot of research on SC models and its modifications has been done \cite{Afanasyev2021, Miao2023, Wei2021, Prasad2018, Guo2019, Smith2021, Kulagin2022, Dull2021}. We adapted these SC models in this paper to fulfil the requirements of modelling of hydrogen chemical reaction.
	
	This paper is organized as follows. In Section \ref{sec:Model}, we introduce the theoretical model, considering both nuclear and electron spins. We also consider the effects of photonic modes $\Omega^s$ and $\Omega^n$ on the quantum evolution and the formation of neutral hydrogen molecule. Some technical details of density matrix and Hamiltonian are included in Section \ref{sec:HamilDensity}. Then we get some results from simulations in Section \ref{sec:Simulations}. Some brief comments on our results and extension to future work in Section \ref{sec:ConcluFuture} close out the paper.

	\section{Theoretical Model}
	\label{sec:Model}
	
	\begin{figure}
		\begin{center}
		\includegraphics[width=0.4\textwidth]{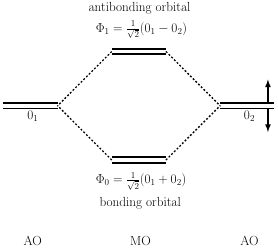}
		\end{center}
		\caption{(online color) The hybridization of orbitals of two hydrogen atoms and the formation of bonding orbital and antibonding orbital.}\label{fig:Hybridization}
	\end{figure}

	The theoretical model, called the association-dissociation model of neutral hydrogen molecule, is detailed in our earlier work \cite{Miao2023}. Each energy level in this model, including atomic and molecular, is divided into two levels: spin up $\uparrow$ and spin down $\uparrow$. According to the Pauli exclusion principle \cite{Pauli1925}, there can only be one electron per level. The excited states of the electron with the spins for the first nucleus is denoted by $|0_1^{\uparrow}\rangle_e$ and $|0_1^{\downarrow}\rangle_e$ (usually simply written as $|0_1\rangle_e$, which can denote both $|0_1^{\uparrow}\rangle_e$ and $|0_1^{\downarrow}\rangle_e$). Then, $|-1_1\rangle_e$ --- ground electron states for the first nucleus. For the second nucleus --- $|0_2\rangle_e$ and $|-1_2\rangle_e$. Hybridization of orbitals is possible only for atomic excited states $|0_{1,2}\rangle_e$. Hybridization of atomic orbitals (AO) and formation of molecular orbitals (MO) are shown in Fig. \ref{fig:Hybridization}, where antibonding orbital and bonding orbital take the following forms, respectively
	\begin{subequations}
		\label{eq:MolStates}
		\begin{align}
			&|\Phi_1\rangle_e=\frac{1}{\sqrt{2}}\left(|0_1\rangle_e-|0_2\rangle_e\right)\label{eq:MolStatePhi1}\\
			&|\Phi_0\rangle_e=\frac{1}{\sqrt{2}}\left(|0_1\rangle_e+|0_2\rangle_e\right)\label{eq:MolStatePhi0}
		\end{align}
	\end{subequations}
	where $|\Phi_1\rangle_e$ are also called molecular excited states, and $|\Phi_1\rangle_e$ --- molecular ground states.

	Each nucleus will form a potential well around itself, and the electrons will be bound in these potential wells. The association reaction of $\mathrm{H}_2$ is described as follows: two electrons in the atomic ground orbital $-1$ with large distance between nuclei, corresponding to different directions of the spin, absorb respectively photon with mode $\Omega^{\uparrow}$ or $\Omega^{\downarrow}$, then they rise to atomic excited orbital $0$. When nuclei gather together in one cavity from different cavities through the quantum tunnelling effect, the potential barrier between the two potential wells decreases, and since the two electrons are in atomic excited orbitals, the atomic orbitals are hybridized into molecular orbitals, and the electrons are released on the molecular excited orbital $\Phi_1$. Then two electrons fleetly release respectively photon with mode $\omega^{\uparrow}$ or $\omega^{\downarrow}$, and fall to molecular ground orbital $\Phi_0$, stable molecule is formed. The dissociation reaction of $\mathrm{H}_2$ is the reverse process of the association reaction, and finally the decomposition of hydrogen molecules is obtained.

	In this paper we adopt the second quantization \cite{Dirac1927, Fock1932}. The entire system's Hilbert space for quantum states is $\mathcal{C}$ and takes the following form
	\begin{equation}
		\label{eq:SpaceC}
		|\Psi\rangle_{\mathcal{C}}=|photon\rangle|electron\rangle|nucleus\rangle
	\end{equation}
	where the quantum state consists of three parts
	\begin{widetext}:
		\begin{subequations}
			\label{eq:Subsystems}
			\begin{align}
				|photon\rangle&=|p_1\rangle_{\omega^{\uparrow}}|p_2\rangle_{\omega^{\downarrow}}|p_3\rangle_{\Omega^{\uparrow}}|p_4\rangle_{\Omega^{\downarrow}}|p_5\rangle_{\Omega^s}\label{eq:PhotonState}\\
			|electron\rangle&=|l_1\rangle_{\substack{at_1\\or_0}}^{\uparrow}|l_2\rangle_{\substack{at_1\\or_0}}^{\downarrow}|l_3\rangle_{\substack{at_1\\or_{-1}}}^{\uparrow}|l_4\rangle_{\substack{at_1\\or_{-1}}}^{\downarrow}|l_5\rangle_{\substack{at_2\\or_0}}^{\uparrow}|l_6\rangle_{\substack{at_2\\or_0}}^{\downarrow}|l_7\rangle_{\substack{at_2\\or_{-1}}}^{\uparrow}|l_8\rangle_{\substack{at_2\\or_{-1}}}^{\downarrow}\label{eq:ElectronState}\\
			|nucleus\rangle&=|k\rangle_n\label{eq:NucleusState}
			\end{align}
		\end{subequations}
	\end{widetext}
	where the numbers of molecule photons with the modes $\omega^{\uparrow}$, $\omega^{\downarrow}$ are $p_1$, $p_2$, respectively; $p_3$, $p_4$ are the numbers of atomic photons with modes $\Omega^{\uparrow}$, $\Omega^{\downarrow}$, respectively; $p_5$ is the number of photons with mode $\Omega^s$, which can excite the electron spin from $\uparrow$ to $\downarrow$ in the atom. $l_{i,i\in\left\{1,2,\cdots,8\right\}}$ describes atom state: $l_i=1$ --- the orbital is occupied by one electron, $l_i=0$ --- the orbital is freed. The state of the nuclei is denoted by $|k\rangle_n$: $k=0$ --- state of nuclei, gathering together in one cavity, $k=1$ --- state of nuclei, scattering in different cavities.

	\subsection{Nuclear and electron spins}
	\label{subsec:NuclearElectronSpin}

	We introduce spin photons with mode $\Omega^s$ in our model, thus transition between $\uparrow$ and $\downarrow$ is allowed. Electron spins must strictly satisfy the Pauli exclusion principle. We stipulate, that independent electron spin transition is allowed if and only if electrons are in atomic excited state. Since this transition will obscure the spin-spin interaction between electron and nucleus (this interaction can only occur when the electron is in the ground state, and is very weak compared to the independent electron spin transition) when the electron is in the ground state. Electron spin transition is also forbidden when electrons are in molecular state corresponding to $|0\rangle_n$, which contravenes Pauli exclusion principle. The stable formation of $\mathrm{H}_2$ is only realized through state, where two electrons with different spins situated in orbital $\Phi_0$.

	Only when the electrons reach the atomic ground state does nuclear spin interact with them. When an electron is in the ground state of an atom and its spin is different from that of the nucleus, they can exchange spins. The symbol for this interaction, known as the spin-spin interaction, is $\sigma_{en,i}$, here $i$ is index of atoms. With the aid of this interaction, the electron with the $\downarrow$ absorbs a photon with mode $\Omega^s$, and the nucleus with the $\uparrow$ emits a photon with mode $\Omega^n$. Now electron spin up and nucleus spin down. In contrast, the nucleus with the $\downarrow$ can also absorb photons with mode $\Omega^n$, and the electron with the $\uparrow$ can also release photons with mode $\Omega^s$.

\begin{figure}
		\begin{center}
		\includegraphics[width=0.5\textwidth]{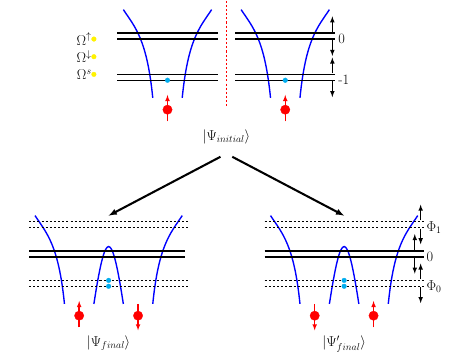}
		\end{center}
		\caption{(online color) The situation with consideration of spin-spin interaction between nucleus and electron. The formation of neutral hydrogen molecule is possible when we put three photons with different modes $\Omega^{\uparrow}$, $\Omega^{\downarrow}$ and $\Omega^s$, and both nuclei are with spin up at the start. Blue and yellow dots represent electrons and photons, respectively. Red up and down arrows represent nuclear spins $\uparrow$ and $\downarrow$, respectively.}\label{fig:Spin-spinInteraction}
	\end{figure}

	The initial state $|\Psi_{initial}\rangle$ for the association process is shown in Fig. \ref{fig:Spin-spinInteraction}, where two electron with $\downarrow$ are in different atoms, and two nuclei with $\uparrow$ can interact with electrons and exchange spins. We put three photons with different modes $\Omega^{\uparrow}$, $\Omega^{\downarrow}$ and $\Omega^s$ at the start. This means that only one of the electrons can complete the spin exchange with the nucleus, because at the initial moment we only have one spin photon with mode $\Omega^s$. Thus, we have two situations of formation of $\mathrm{H}_2$:
	\begin{itemize}
		\item the first nucleus with $\uparrow$ and the second nucleus with $\downarrow$, denoted by $|\Psi_{final}\rangle$;
		\item the first nucleus with $\downarrow$ and the second nucleus with $\uparrow$, denoted by $|\Psi_{final}'\rangle$.
	\end{itemize}

	Stable hydrogen molecule is defined as follows
	\begin{equation}
		\label{eq:HydrogenState}
		|\mathrm{H}_2\rangle=c_0|\Psi_{final}\rangle+c_1|\Psi_{final}'\rangle
	\end{equation}
	where $c_0$, $c_1$ are normalization factors.

	Due to the introduction of nuclear spin, we need to introduce nuclear spin photon with mode $\Omega^n$ and consider the spin state of the two nuclei. Thus, the definition of quantum state space must be rewritten. Above all, Eq. (\ref{eq:PhotonState}) is expended as follows
	\begin{equation}
		\label{eq:NewPhotonState}
		|photon\rangle=|p_1\rangle_{\omega^{\uparrow}}|p_2\rangle_{\omega^{\downarrow}}|p_3\rangle_{\Omega^{\uparrow}}|p_4\rangle_{\Omega^{\downarrow}}|p_5\rangle_{\Omega^s}|p_6\rangle_{\Omega^n}
	\end{equation}
	where $p_6$ is the number of photons with mode $\Omega^n$, which can excite the nuclear spin from $\downarrow$ to $\uparrow$ in the atom. Analogously, Eq. (\ref{eq:NucleusState}) is expended as follows
	\begin{equation}
		\label{eq:NewNucleusState}
		|nucleus\rangle=|k\rangle_n|k_1\rangle_{n_1}|k_2\rangle_{n_2}
	\end{equation}
	where $k_{i,i\in\left\{1,2\right\}}$ describes nuclear spin of the first or second atom. $k_i=1$ --- nucleus with $\uparrow$, $k_i=0$ --- nucleus with $\downarrow$.

	Spin-spin interaction between nucleus and electron with slight intensity $g_{en}$ is usually ignored. However, experiments indicate that when we introduce spin-spin interaction, molecular hydrogen occurs in two isomeric forms: one with its two proton nuclear spins aligned parallel --- orthohydrogen, the other with its two proton spins aligned antiparallel --- parahydrogen. The spin-spin interaction is also called hyperfine.
	
	\subsection{Thermally stationary state}
	\label{subsec:Thermally}
	
	We define the stationary state of a field with temperature $T$ as a mixed state with a Gibbs distribution of Fock components
	\begin{equation}
		\label{eq:Photon_gibbs}
		{\cal G}\left(T\right)_f=c\sum\limits_{p=0}^\infty exp\left(-\frac{\hbar\omega_c p}{KT}\right)|p\rangle\langle p|
	\end{equation} 
where $K$ is the Boltzmann constant, $c$ is the normalization factor, $p$ is the number of photons, $\omega_c$ is the photonic mode. We introduce the notation $\gamma_{k'}/\gamma_{k}=\mu$, where $\gamma_{k}$ denotes the total spontaneous emission rate for photon from cavity to external environment and $\gamma_{k'}$ denotes the total spontaneous influx rate for photon from external environment into cavity. The state ${\cal G}\left(T\right)_f$ will then exist only at $\mu<1$, because otherwise the temperature will be infinitely large and the state ${\cal G}\left(T\right)_f$ will be non-normalizable. The population of the photonic Fock state $|p\rangle$ at temperature $T$ is proportional to $exp\left(-\frac{\hbar\omega_c}{KT}\right)$. In our model, we assume
	\begin{equation}
		\label{eq:PopulationFock}
		\mu=exp\left(-\frac{\hbar\omega_c}{KT}\right)
	\end{equation}
	from where $T=\frac{\hbar\omega_c}{K\ln\left(1/\mu\right)}$.

	The following theorem takes place \cite{Kulagin2018}:

	The thermally stationary state of atoms and fields at temperature $T$ has the form $\rho_{state}=\rho_{ph}\otimes\rho_{at}$, where $\rho_{ph}$ is the state of the photon and $\rho_{at}$ is the state of the atom.
	
	\section{Hamiltonian and density matrix}
	\label{sec:HamilDensity}
	
	The quantum master equation (QME) in the Markovian approximation for the density operator $\rho$ of the system takes the following form
	\begin{equation}
		\label{eq:QME}
		i\hbar\dot{\rho}=\left[H,\rho\right]+iL\left(\rho\right)
	\end{equation}
	where $L\left(\rho\right)$ is as follows
	\begin{equation}
		\label{eq:LindbladOperator}
		\begin{aligned}
			L\left(\rho\right)&=\sum_{k\in \mathcal{K}}L_k\left(\rho\right)+\sum_{k'\in \mathcal{K}'}L_{k'}\left(\rho\right)\\
			&=\sum_{k\in \mathcal{K}}\gamma_k\left(A_k\rho A_k^{\dag}-\frac{1}{2}\left\{\rho, A_k^{\dag}A_k\right\}\right)\\
			&+\sum_{k'\in \mathcal{K}'}\gamma_{k'}\left(A_k^{\dag}\rho A_k-\frac{1}{2}\left\{\rho, A_kA_k^{\dag}\right\}\right)
		\end{aligned}
	\end{equation}
	where $\mathcal{K}$ is a graph of the potential photon dissipations between the states that are permitted. The edges and vertices of $\mathcal{K}$ represent the permitted dissipations and the states, respectively. $\mathcal{K}'$ is a graph of the potential photon influxes between the states that are permitted. $L_k\left(\rho\right)$ ($L_{k'}\left(\rho\right)$) is the standard dissipation (influx) superoperator corresponding to the jump operator $A_k$ ($A_{k'}$), and the term $\gamma_k$ ($\gamma_{k'}$) refers to the overall spontaneous emission (influx) rate for photons for $k\in \mathcal{K}$ ($k'\in\mathcal{K}'$).

	The coupled-system Hamiltonian in Eq. \eqref{eq:QME} is expressed by the total energy operator
	\begin{equation}
		\label{eq:Hamil}
		H=H_{\mathcal{A}}+H_{\mathcal{D}}+H_{tun}+H_{spin-flip}+H_{spin-spin}
	\end{equation}
	where $H_{tun}$ denotes the quantum tunnelling effect between $H_{\mathcal{A}}$ and $H_{\mathcal{D}}$, which are the associative and dissociative Hamiltonians, respectively. $H_{spin-flip}$ describes the electron spin transition (spin-flip) and $H_{spin-spin}$ denotes the spin-spin interaction between nucleus and electron.
	
	Rotating wave approximation (RWA) \cite{Wu2007} is taken into account
	\begin{equation}
		\label{eq:RWACondition}
		\frac{g}{\hbar\omega_c}\approx\frac{g}{\hbar\omega_n}\ll 1
	\end{equation}
where $\omega_c$ stands for cavity frequency and $\omega_n$ for transition frequency. We presume that $\omega_c=\omega_n$.
	
	We will directly quote and transform the definitions of $H_{\mathcal{A}},\ H_{\mathcal{D}},\ H_{tun},\ H_{spin-flip}$ from our earlier paper \cite{Miao2023}. Thus, $H_{\mathcal{A}}$ has following form
	\begin{equation}
		\label{eq:HamilA}
		H_{\mathcal{A}}=\left(H_{\mathcal{A},field}+H_{\mathcal{A},mol}+H_{\mathcal{A},int}\right)\sigma_n\sigma_n^{\dag}
	\end{equation}
	where $\sigma_n\sigma_n^{\dag}$ verifies that nuclei are close. And
	\begin{subequations}
		\label{eq:HamilADetail}
		\begin{align}
			&H_{\mathcal{A},field}=\hbar\omega^{\uparrow}a_{\omega^{\uparrow}}^{\dag}a_{\omega^{\uparrow}}+\hbar\omega^{\downarrow}a_{\omega^{\downarrow}}^{\dag}a_{\omega^{\downarrow}}\label{eq:HamilAField}\\
			&H_{\mathcal{A},mol}=\hbar\omega^{\uparrow}\sigma_{\omega^{\uparrow}}^{\dag}\sigma_{\omega^{\uparrow}}+\hbar\omega^{\downarrow}\sigma_{\omega^{\downarrow}}^{\dag}\sigma_{\omega^{\downarrow}}\label{eq:HamilAMol}\\
			&H_{\mathcal{A},int}=g_{\omega^{\uparrow}}\left(a_{\omega^{\uparrow}}^{\dag}\sigma_{\omega^{\uparrow}}+a_{\omega^{\uparrow}}\sigma_{\omega^{\uparrow}}^{\dag}\right)\\
			&+g_{\omega^{\downarrow}}\left(a_{\omega^{\downarrow}}^{\dag}\sigma_{\omega^{\downarrow}}+a_{\omega^{\downarrow}}\sigma_{\omega^{\downarrow}}^{\dag}\right)\label{eq:HamilAInt}
		\end{align}
	\end{subequations}
	where $\hbar$ is the reduced Planck constant or Dirac constant. $H_{\mathcal{A},field}$ is the photon energy operator, $H_{\mathcal{A},mol}$ is the molecule energy operator, $H_{\mathcal{A},int}$ is the molecule-photon interaction operator. $g_{\omega}$ is the coupling strength between the photon mode $\omega$ (with annihilation and creation operators $a_{\omega}$ and $a_{\omega}^{\dag}$, respectively) and the electrons (with excitation and relaxation operators $\sigma_{\omega}^{\dag}$ and $\sigma_{\omega}$, respectively). 

	Then $H_{\mathcal{D}}$ is described in following form
	\begin{equation}
		\label{eq:HamilD}
		H_{\mathcal{D}}=\left(H_{\mathcal{D},field}+H_{\mathcal{D},mol}+H_{\mathcal{D},int}\right)\sigma_n^{\dag}\sigma_n
	\end{equation}
where $\sigma_n^{\dag}\sigma_n$ verifies that nuclei are far away. And
	\begin{subequations}
		\label{eq:HamilDDetail}
		\begin{align}
			&H_{\mathcal{D},field}=\hbar\Omega^{\uparrow}a_{\Omega^{\uparrow}}^{\dag}a_{\Omega^{\uparrow}}+\hbar\Omega^{\downarrow}a_{\Omega^{\downarrow}}^{\dag}a_{\Omega^{\downarrow}}\label{eq:HamilDField}\\
			&H_{\mathcal{D},at}=\sum_{i=1,2}\left(\hbar\Omega^{\uparrow}\sigma_{\Omega^{\uparrow},i}^{\dag}\sigma_{\Omega^{\uparrow},i}+\hbar\Omega^{\downarrow}\sigma_{\Omega^{\downarrow},i}^{\dag}\sigma_{\Omega^{\downarrow},i}\right)\label{eq:HamilDAt}\\
			&H_{\mathcal{D},int}=\sum_{i=1,2}\left\{g_{\Omega^{\uparrow}}\left(a_{\Omega^{\uparrow}}^{\dag}\sigma_{\Omega^{\uparrow},i}+a_{\Omega^{\uparrow}}\sigma_{\Omega^{\uparrow},i}^{\dag}\right)\right.\\
			&\left.+g_{\Omega^{\downarrow}}\left(a_{\Omega^{\downarrow}}^{\dag}\sigma_{\Omega^{\downarrow},i}+a_{\Omega^{\downarrow}}\sigma_{\Omega^{\downarrow},i}^{\dag}\right)\right\}\label{eq:HamilDInt}
		\end{align}
	\end{subequations}
where $H_{\mathcal{D},field}$ is the photon energy operator, $H_{\mathcal{D},at}$ is the atom energy operator, $H_{\mathcal{D},int}$ is atom-photon interaction operator. $g_{\Omega}$ is the coupling strength between the photon mode $\Omega$ (with annihilation and creation operators $a_{\Omega}$ and $a_{\Omega}^{\dag}$, respectively) and the electrons in the atom (with excitation and relaxation operators $\sigma_{\Omega,i}^{\dag}$ and $\sigma_{\Omega,i}$, respectively, here $i$ denotes index of atoms).

	$H_{tun}$ describe the hybridization and de-hybridization, realized by quantum tunnelling effect, it takes the form
	\begin{equation}
		\label{eq:HamilTDetail}
		\begin{aligned}
			H_{tun}&=\zeta_2\sigma_{\omega^{\uparrow}}^{\dag}\sigma_{\omega^{\uparrow}}\sigma_{\omega^{\downarrow}}^{\dag}\sigma_{\omega^{\downarrow}}\left(\sigma_n^{\dag}+\sigma_n\right)\\
			&+\zeta_1\sigma_{\omega^{\uparrow}}\sigma_{\omega^{\uparrow}}^{\dag}\sigma_{\omega^{\downarrow}}^{\dag}\sigma_{\omega^{\downarrow}}\left(\sigma_n^{\dag}+\sigma_n\right)\\
			&+\zeta_1\sigma_{\omega^{\uparrow}}^{\dag}\sigma_{\omega^{\uparrow}}\sigma_{\omega^{\downarrow}}\sigma_{\omega^{\downarrow}}^{\dag}\left(\sigma_n^{\dag}+\sigma_n\right)\\
			&+\zeta_0\sigma_{\omega^{\uparrow}}\sigma_{\omega^{\uparrow}}^{\dag}\sigma_{\omega^{\downarrow}}\sigma_{\omega^{\downarrow}}^{\dag}\left(\sigma_n^{\dag}+\sigma_n\right)
		\end{aligned}
	\end{equation}
	where $\sigma_{\omega^{\uparrow}}^{\dag}\sigma_{\omega^{\uparrow}}\sigma_{\omega^{\downarrow}}^{\dag}\sigma_{\omega^{\downarrow}}$ verifies that two electrons with different spins are at orbital $\Phi_1$ with large tunnelling intensity $\zeta_2$; $\sigma_{\omega^{\uparrow}}\sigma_{\omega^{\uparrow}}^{\dag}\sigma_{\omega^{\downarrow}}^{\dag}\sigma_{\omega^{\downarrow}}$ verifies that electron with $\uparrow$ is at orbital $\Phi_0$ and electron with $\downarrow$ is at orbital $\Phi_1$, with low tunnelling intensity $\zeta_1$; $\sigma_{\omega^{\uparrow}}^{\dag}\sigma_{\omega^{\uparrow}}\sigma_{\omega^{\downarrow}}\sigma_{\omega^{\downarrow}}^{\dag}$ verifies that electron with $\uparrow$ is at orbital $\Phi_1$ and electron with $\downarrow$ is at orbital $\Phi_0$, with low tunnelling intensity $\zeta_1$; $\sigma_{\omega^{\uparrow}}\sigma_{\omega^{\uparrow}}^{\dag}\sigma_{\omega^{\downarrow}}\sigma_{\omega^{\downarrow}}^{\dag}$ verifies that two electrons with different spins are at orbital $\Phi_0$ with tunnelling intensity $\zeta_0$, which equal to $0$.
	
	We assume that the electron spin transition only occurs when the electron is in the atomic excited state, and we only consider the spin-spin interaction of the electron with the nucleus when the electron is in the atomic ground state. Thus, $H_{spin-flip}$ takes the form
	\begin{widetext}
		\begin{equation}
			\label{eq:HamilSpinflip}
			H_{spin-flip}=\sum_{i=1,2}\left\{\left(\sigma_{\Omega^{\uparrow},i}^{\dag}\sigma_{\Omega^{\uparrow},i}+\sigma_{\Omega^{\downarrow},i}^{\dag}\sigma_{\Omega^{\downarrow},i}\right)\left[\hbar\Omega^sa_{\Omega^s}^{\dag}a_{\Omega^s}+\hbar\Omega^s\sigma_{\Omega^s,i}^{\dag}\sigma_{\Omega^s,i}+g_{\Omega^s}\left(a_{\Omega^s}^{\dag}\sigma_{\Omega^s,i}+a_{\Omega^s}\sigma_{\Omega^s,i}^{\dag}\right)\right]\right\}
		\end{equation}
	\end{widetext}
	where $\sigma_{\Omega^{\uparrow},i}^{\dag}\sigma_{\Omega^{\uparrow},i}+\sigma_{\Omega^{\downarrow},i}^{\dag}\sigma_{\Omega^{\downarrow},i}$ verifies that electron is in the atomic excited state. $i$ denotes index of atoms.
	
	$H_{spin-spin}$ tasks the form
	\begin{widetext}
		\begin{equation}
			\label{eq:HamilSpinSpin}
			\begin{aligned}
				H_{spin-spin}&=\sum_{i=1,2}\left\{\left(\sigma_{\Omega^{\uparrow},i}\sigma_{\Omega^{\uparrow},i}^{\dag}+\sigma_{\Omega^{\downarrow},i}\sigma_{\Omega^{\downarrow},i}^{\dag}\right)\left[\hbar\Omega^sa_{\Omega^s}^{\dag}a_{\Omega^s}+\hbar\Omega^s\sigma_{\Omega^s,i}^{\dag}\sigma_{\Omega^s,i}\right.\right.\\
				&\left.\left.+\hbar\Omega^na_{\Omega^n}^{\dag}a_{\Omega^n}+\hbar\Omega^n\sigma_{\Omega^n,i}^{\dag}\sigma_{\Omega^n,i}+g_{en}\left(\sigma_{en,i}+\sigma_{en,i}^{\dag}\right)\right]\right\}
			\end{aligned}
		\end{equation}
	\end{widetext}
	where $\sigma_{\Omega^{\uparrow},i}\sigma_{\Omega^{\uparrow},i}^{\dag}+\sigma_{\Omega^{\downarrow},i}\sigma_{\Omega^{\downarrow},i}^{\dag}$ verifies that electron is in the atomic ground state. And $\sigma_{en,i}$ takes the form
	\begin{equation}
		\label{eq:OperatorElectronNucleus}
		\sigma_{en,i} = a_{\Omega^s}\sigma_{\Omega^s,i}^{\dag}a_{\Omega^n}^{\dag}\sigma_{\Omega^n,i}
	\end{equation}
	and $\sigma_{en,i}^{\dag}$ is its hermitian conjugate operator.

	On a p-photons state, the photon annihilation and creation operators $a$ and $a^{\dag}$ are described as
	\begin{equation}
		\label{eq:PhotonOperators}
		\begin{aligned}
			&if\ p>0,\ \left\{
				\begin{aligned}
				&a|p\rangle=\sqrt{p}|p-1\rangle,\\
				&a^{\dag}|p\rangle=\sqrt{p+1}|p+1\rangle,
				\end{aligned}
				\right
				.\\
			&if\ p=0,\ \left \{
				\begin{aligned}
					&a|0\rangle=0,\\
					&a^{\dag}|0\rangle=|1\rangle.
				\end{aligned}
				\right
				.\\
		\end{aligned}
	\end{equation}
	Operators $a_{\omega^{\uparrow}}$, $a_{\omega^{\downarrow}}$, $a_{\Omega^{\uparrow}}$, $a_{\Omega^{\downarrow}}$, $a_{\Omega^s}$, $a_{\Omega^n}$ and their hermitian conjugate operators all obey the rules in \eqref{eq:PhotonOperators}.
	
	The interaction of molecule with the electromagnetic field of the cavity, emitting or absorbing photon with mode $\omega^{\uparrow,\downarrow}$, is described as
	\begin{equation}
		\label{eq:InteractionMolecule}
		\begin{aligned}
			&\sigma_{\omega^{\uparrow}}|1\rangle_{\Phi_1}^{\uparrow}|0\rangle_{\Phi_0}^{\uparrow}=|0\rangle_{\Phi_1}^{\uparrow}|1\rangle_{\Phi_0}^{\uparrow},\\
			&\sigma_{\omega^{\uparrow}}^{\dag}|0\rangle_{\Phi_1}^{\uparrow}|1\rangle_{\Phi_0}^{\uparrow}=|1\rangle_{\Phi_1}^{\uparrow}|0\rangle_{\Phi_0}^{\uparrow},\\
			&\sigma_{\omega^{\downarrow}}|1\rangle_{\Phi_1}^{\downarrow}|0\rangle_{\Phi_0}^{\downarrow}=|0\rangle_{\Phi_1}^{\downarrow}|1\rangle_{\Phi_0}^{\downarrow},\\
			&\sigma_{\omega^{\downarrow}}^{\dag}|0\rangle_{\Phi_1}^{\downarrow}|1\rangle_{\Phi_0}^{\downarrow}=|1\rangle_{\Phi_1}^{\downarrow}|0\rangle_{\Phi_0}^{\downarrow}.
		\end{aligned}
	\end{equation}

	The interaction of atom with the electromagnetic field of the cavity, emitting or absorbing photon with mode $\Omega^{\uparrow,\downarrow}$, is described as
	\begin{equation}
		\label{eq:InteractionAtom}
		\begin{aligned}
			&\sigma_{\Omega^{\uparrow},i}|1\rangle_{\substack{at_i\\or_0}}^{\uparrow}|0\rangle_{\substack{at_i\\or_{-1}}}^{\uparrow}=|0\rangle_{\substack{at_i\\or_0}}^{\uparrow}|1\rangle_{\substack{at_i\\or_{-1}}}^{\uparrow},\\
			&\sigma_{\Omega^{\uparrow},i}^{\dag}|0\rangle_{\substack{at_i\\or_0}}^{\uparrow}|1\rangle_{\substack{at_i\\or_{-1}}}^{\uparrow}=|1\rangle_{\substack{at_i\\or_0}}^{\uparrow}|0\rangle_{\substack{at_i\\or_{-1}}}^{\uparrow},\\
			&\sigma_{\Omega^{\downarrow},i}|1\rangle_{\substack{at_i\\or_0}}^{\downarrow}|0\rangle_{\substack{at_i\\or_{-1}}}^{\downarrow}=|0\rangle_{\substack{at_i\\or_0}}^{\downarrow}|1\rangle_{\substack{at_i\\or_{-1}}}^{\downarrow},\\
			&\sigma_{\Omega^{\downarrow},i}^{\dag}|0\rangle_{\substack{at_i\\or_0}}^{\downarrow}|1\rangle_{\substack{at_i\\or_{-1}}}^{\downarrow}=|1\rangle_{\substack{at_i\\or_0}}^{\downarrow}|0\rangle_{\substack{at_i\\or_{-1}}}^{\downarrow}.
		\end{aligned}
	\end{equation}
	
	The nuclei's tunnelling operators have following form
	\begin{equation}
		\label{eq:TunnellingOperators}
		\begin{aligned}
			&\sigma_n|1\rangle_n=|0\rangle_n,\\
			&\sigma_n^{\dag}|0\rangle_n=|1\rangle_n.
		\end{aligned}
	\end{equation}

	And the interaction of atom with the electromagnetic field of the cavity, emitting or absorbing photon with mode $\Omega^s$ and causing electron spin-flip, is described as
	\begin{equation}
		\label{eq:InteractionAtomSpin}
		\begin{aligned}
			&\sigma_{\Omega^s,i}|1\rangle_{\substack{at_i,\\or_0}}^{\uparrow}|0\rangle_{\substack{at_i,\\or_0}}^{\downarrow}=|0\rangle_{\substack{at_i,\\or_0}}^{\uparrow}|1\rangle_{\substack{at_i,\\or_0}}^{\downarrow},\\
			&\sigma_{\Omega^s,i}^{\dag}|0\rangle_{\substack{at_i,\\or_0}}^{\uparrow}|1\rangle_{\substack{at_i,\\or_0}}^{\downarrow}=|1\rangle_{\substack{at_i,\\or_0}}^{\uparrow}|0\rangle_{\substack{at_i,\\or_0}}^{\downarrow},\\
			&\sigma_{\Omega^s,i}|1\rangle_{\substack{at_i,\\or_{-1}}}^{\uparrow}|0\rangle_{\substack{at_i,\\or_{-1}}}^{\downarrow}=|0\rangle_{\substack{at_i,\\or_{-1}}}^{\uparrow}|1\rangle_{\substack{at_i,\\or_{-1}}}^{\downarrow},\\
			&\sigma_{\Omega^s,i}^{\dag}|0\rangle_{\substack{at_i,\\or_{-1}}}^{\uparrow}|1\rangle_{\substack{at_i,\\or_{-1}}}^{\downarrow}=|1\rangle_{\substack{at_i,\\or_{-1}}}^{\uparrow}|0\rangle_{\substack{at_i,\\or_{-1}}}^{\downarrow}.
		\end{aligned}
	\end{equation}
	
	And the interaction of nucleus with the electromagnetic field of the cavity, emitting or absorbing photon with mode $\Omega^n$ and causing nuclear spin-flip, is described as
	\begin{equation}
		\label{eq:InteractionNucleusSpin}
		\begin{aligned}
			&\sigma_{\Omega^n,i}|1\rangle_{n_i}=|0\rangle_{n_i},\\
			&\sigma_{\Omega^n,i}^{\dag}|0\rangle_{n_i}=|1\rangle_{n_i}.
		\end{aligned}
	\end{equation}

	\section{Simulations and results}
	\label{sec:Simulations}
	
	\begin{figure}
		\begin{center}
		\includegraphics[width=0.5\textwidth]{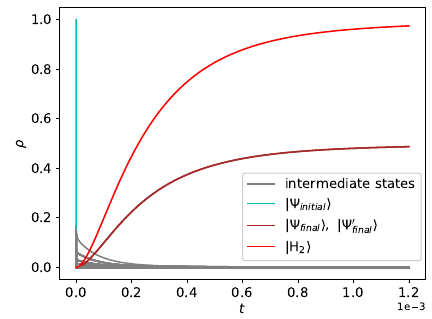}
		\end{center}
		\caption{(online color) The evolution with consideration of electron spin transition and spin-spin interaction between electrons and nuclei. Probability of state $|\Psi_{initial}\rangle$ is denoted by cyan solid curve, and probability of states $|\Psi_{final}\rangle$ and $|\Psi_{final}'\rangle$ is denoted by brown solid curve, the curve of time-dependent probability of state $|\mathrm{H}_2\rangle$ is denoted by red solid curve, and other intermediate states are in gray.}\label{fig:Formation}
	\end{figure}
	
	Now we introduce the numerical method to simulate the evolution of system. The solution $\rho\left(t\right)$ in Eq. \eqref{eq:QME} may be approximately found as a sequence of two steps. In the first step we make one step in the solution of the unitary part of Eq. \eqref{eq:QME}
	\begin{equation}
		\label{eq:UnitaryPart}
		\tilde{\rho}\left(t+dt\right)=exp\left(-\frac{i}{\hbar}Hdt\right)\rho\left(t\right)exp\left(\frac{i}{\hbar}Hdt\right)
	\end{equation}
and in the second step, make one step in the solution of Eq. \eqref{eq:QME} with the commutator removed
	\begin{equation}
		\label{eq:Solution}
		\rho\left(t+dt\right)=\tilde{\rho}\left(t+dt\right)+\frac{1}{\hbar}L\left(\tilde{\rho}\left(t+dt\right)\right)dt
	\end{equation}
	
	In simulations:

	$\hbar=1$, $\Omega^{\uparrow}=\Omega^{\downarrow}=10^{10}$, $\omega^{\uparrow}=\omega^{\downarrow}=5*10^9$, $\Omega^s=10^9$, $\Omega^n=10^8$;

	$g_{\Omega^{\uparrow}}=g_{\Omega^{\uparrow}}=10^8$, $g_{\omega^{\uparrow}}=g_{\omega^{\uparrow}}=5*10^7$, $g_{\Omega^s}=10^7$, $g_{en}=10^6$, $\zeta_2=10^9$, $\zeta_1=10^7$, $\zeta_0=0$.

	We consider the leakage of all types of photon in Markovian open systems, and its dissipative rate all are equal:

	$\gamma_{\Omega^{\uparrow}}=\gamma_{\Omega^{\downarrow}}=\gamma_{\omega^{\uparrow}}=\gamma_{\omega^{\downarrow}}=\gamma_{\Omega^s}=\gamma_{\Omega^n}=10^7$.

	\subsection{Formation of neutral hydrogen molecule}
	\label{subsec:Formation}
	
	 In Subsection \ref{subsec:NuclearElectronSpin}, we introduce feeble spin-spin interaction between electrons and nuclei. Initial state is $|\Psi_{initial}\rangle$, described in Fig. \ref{fig:Spin-spinInteraction}, where two electrons with $\downarrow$ are both in atomic ground orbital as above, and two nuclei are with $\uparrow$. Spin-spin interaction is permissible, and it only happens when electron is in atomic ground orbital, which is close to nucleus. Comparing to independent electron spin transition strength $g_{\Omega^s}$, strength of spin-spin interaction between nucleus and electron $g_{en}$ is extremely slight. Thus, we provisionally neglect independent electron spin transition in ground orbitals in order to study the effect of spin-spin interaction on formation of neutral hydrogen molecule.

	We assume that $\mu_{\Omega^{\uparrow}}=\mu_{\Omega^{\downarrow}}=\mu_{\Omega^s}=\mu_{\Omega^n}=0.5$ and $\mu_{\omega^{\uparrow}}=\mu_{\omega^{\downarrow}}=0$. This means that photons with modes $\Omega^{\uparrow}$, $\Omega^{\downarrow}$, $\Omega^s$, $\Omega^n$ will be continuously injected into the system, while another photons with $\omega^{\uparrow}$, $\omega^{\downarrow}$ will not be replenished.

	According to numerical results in Fig. \ref{fig:Formation}, we found brown solid curve representing $|\Psi_{final}\rangle$ and $|\Psi_{final}'\rangle$ rises and reaches $0.5$ at the end. $|\Psi_{final}\rangle$ and $|\Psi_{final}'\rangle$ are described in Fig. \ref{fig:Spin-spinInteraction}, where two electrons with different spins are fastened in molecular ground orbital, corresponding to $|0\rangle_n|1\rangle_{n_1}|0\rangle_{n_2}$ and $|0\rangle_n|0\rangle_{n_1}|1\rangle_{n_2}$, respectively. It also means that formation of $\mathrm{H}_2$ ($|\mathrm{H}_2\rangle=c_0|\Psi_{final}\rangle+c_1|\Psi_{final}'\rangle$) is achieved (red solid curve also rises and reaches $1$) and free hydrogen atoms are no longer in existence. Thus, we can say that formation of neutral hydrogen molecule is possible.

	\subsection{Effect of electron spin}
	\label{subsec:EffectElectronSpin}
	
	\begin{figure*}
		\begin{center}
		\includegraphics[width=1\textwidth]{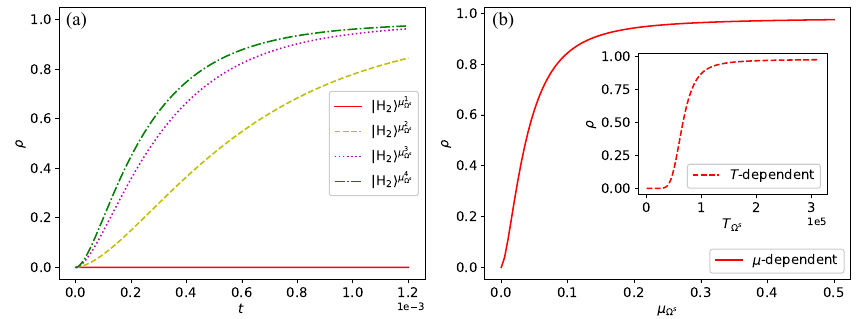}
		\end{center}
		\caption{(online color) Effect of photonic mode $\Omega^s$. In {\bf (a)}, curves of $|\mathrm{H}_2\rangle$ are corresponding to $\mu_{\Omega^s}^1$ (red solid), $\mu_{\Omega^s}^2$ (yellow dashed), $\mu_{\Omega^s}^3$ (magenta dotted) and $\mu_{\Omega^s}^4$ (green dash-dotted), respectively. In {\bf (b)}, red solid curve represent $|\mathrm{H}_2\rangle$, when time of evolution reaches $0.0012s$, with the increases of $\mu_{\Omega^s}$ from $0$ to $0.5$. Red dashed curve in inserted figure represents the $T$-dependent probability of $|\mathrm{H}_2\rangle$ when time reaches $0.0012s$.}\label{fig:EffectElectronSpin}
	\end{figure*}
	
	\begin{figure*}
		\begin{center}
		\includegraphics[width=1\textwidth]{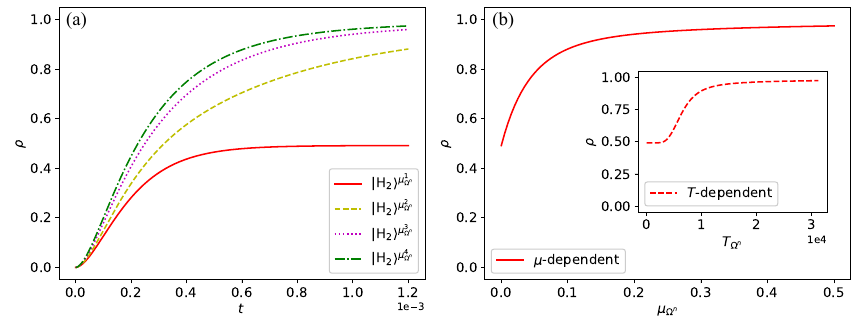}
		\end{center}
		\caption{(online color) Effect of photonic mode $\Omega^n$. In {\bf (a)}, curves of $|\mathrm{H}_2\rangle$ are corresponding to $\mu_{\Omega^n}^1$ (red solid), $\mu_{\Omega^n}^2$ (yellow dashed), $\mu_{\Omega^n}^3$ (magenta dotted) and $\mu_{\Omega^n}^4$ (green dash-dotted), respectively. In {\bf (b)}, red solid curve represent $|\mathrm{H}_2\rangle$, when time of evolution reaches $0.0012s$, with the increases of $\mu_{\Omega^n}$ from $0$ to $0.5$. Red dashed curve in inserted figure represents the $T$-dependent probability of $|\mathrm{H}_2\rangle$ when time reaches $0.0012s$.}\label{fig:EffectNuclearSpin}
	\end{figure*}

	Now we investigate the effect of photonic mode $\Omega^s$ on the evolution and the formation of neutral hydrogen molecule.
	
	We assume that $\mu_{\Omega^{\uparrow}}=\mu_{\Omega^{\downarrow}}=\mu_{\Omega^n}=0.5$ and $\mu_{\omega^{\uparrow}}=\mu_{\omega^{\downarrow}}=0$.

	In Fig. \ref{fig:EffectElectronSpin} \textbf{(a)}, we chose four instances that vary in various $\mu_{\Omega^s}$: $\mu_{\Omega^s}^1=0$, $\mu_{\Omega^s}^2=0.1$, $\mu_{\Omega^s}^3=0.3$, $\mu_{\Omega^s}^4=0.5$. We discovered that neutral hydrogen molecule forms more quickly the higher the $\mu_{\Omega^s}$. The circumstance where $\mu_{\Omega^s}^1=0$ (in this case, $T_{\Omega^s}^1=0K$) occurs is where formation moves the slowest, indicated by red solid curve. The fastest formation occurs when $\mu_{\Omega^s}^4=0.5$, indicated by green dashed-dotted curve. The probability of the $|\mathrm{H}_2\rangle$ never approaches $1$ when the $\mu_{\Omega^s}$ is equal to $0$. However, once $\mu_{\Omega^s}$ is bigger than $0$, the probability of $|\mathrm{H}_2\rangle$ will reach $1$ as long as the duration is long enough. Atomic photons are continuously reintroduced back into the system since molecular photons are not regenerated. As a result, the whole system will gradually change to create a stable molecular state.

	We now raise $\mu_{\Omega^s}$ from $0$ to $0.5$. In each case we take the value of state $|\mathrm{H}_2\rangle$ when the time of evolution is $0.0012s$. We can intuitively perceive the trend of $|\mathrm{H}_2\rangle$ with the growth of $\mu_{\Omega^s}$ in Fig. \ref{fig:EffectElectronSpin} \textbf{(b)}. Probability of $|\mathrm{H}_2\rangle$ is close to 0 when $\mu_{\Omega^s}$ is near to $0$. It begins to increase as the $\mu_{\Omega^s}$ rises, then it reaches a top, which is close to $1$. From the inserted figure in Fig. \ref{fig:EffectElectronSpin} \textbf{(b)}, we can see that the $T$-dependent curve of probability has the same trend as the $\mu$-dependent curve, but there is a hysteresis near $0K$.
	
	\subsection{Effect of nuclear spin}
	\label{subsec:EffectNuclearSpin}
	
	Then we investigate the influence of photonic modes $\Omega^n$ to the evolution and the formation of neutral hydrogen molecule.
	
	We assume that $\mu_{\Omega^{\uparrow}}=\mu_{\Omega^{\downarrow}}=\mu_{\Omega^s}=0.5$ and $\mu_{\omega^{\uparrow}}=\mu_{\omega^{\downarrow}}=0$.

	In Fig. \ref{fig:EffectNuclearSpin} \text{(a)}, we chose four instances that vary in various $\mu_{\Omega^n}$: $\mu_{\Omega^n}^1=0$, $\mu_{\Omega^n}^2=0.1$, $\mu_{\Omega^n}^3=0.3$, $\mu_{\Omega^n}^4=0.5$. We discovered that neutral hydrogen molecule forms more quickly the higher the $\mu_{\Omega^n}$. However, compared with the $\Omega^s$, the promoting effect of the $\Omega^n$ on neutral hydrogen molecular formation is not so great due to the weaker spin-spin interaction. The circumstance where $\mu_{\Omega^n}^1=0$ (in this case, $T_{\Omega^n}^1=0K$) occurs is where formation moves the slowest, indicated by red solid curve. But probability of $|\mathrm{H}_2\rangle$ is much higher than the case where $\mu_{\Omega^n}$ is equal to $0$. This is because in the initial state, we have only one electron spin photon, but there are two nuclei both with $\uparrow$, which means that at most two nuclear spin photons will be released. The fastest formation occurs when $\mu_{\Omega^n}^4=0.5$, indicated by green dashed-dotted curve. The probability of the $|\mathrm{H}_2\rangle$ never approaches $1$ when the $\mu_{\Omega^n}$ is equal to $0$. However, once $\mu_{\Omega^n}$ is bigger than $0$, the probability of $|\mathrm{H}_2\rangle$ will reach $1$ as long as the duration is long enough.

	We now raise $\mu_{\Omega^n}$ from $0$ to $0.5$. In each case we take the value of state $|\mathrm{H}_2\rangle$ when the time of evolution is $0.0012s$. We can intuitively perceive the trend of $|\mathrm{H}_2\rangle$ with the growth of $\mu_{\Omega^n}$ in Fig. \ref{fig:EffectNuclearSpin} \textbf{(b)}. Probability of $|\mathrm{H}_2\rangle$ is close to $0.5$ when $\mu_{\Omega^n}$ is near to $0$. It begins to increase as the $\mu_{\Omega^n}$ rises, then it reaches a top, which is close to $1$. For photonic mode $\Omega^n$, the $T$-dependent curve of probability has a hysteresis, too.

	\section{Concluding discussion and future work} 
	\label{sec:ConcluFuture}
	
	In this paper, we simulate the neutral hydrogen molecule formation in the association-dissociation model of neutral hydrogen molecule. We introduce spin-spin interaction into the system and derived some analytical results of it:
	
	Above all, we studied spin-spin interaction between electrons and nuclei in Subsection \ref{subsec:Formation}. In this part, we investigated the formation of hydrogen molecule. Then the effects of temperature variation of $\Omega^s$ and $\Omega^n$ on the formation of neutral hydrogen molecule is obtained in Subsection \ref{subsec:EffectElectronSpin} and Subsection \ref{subsec:EffectNuclearSpin}: the higher temperature, the faster process of neutral hydrogen molecule formation. We have established the adequacy of our model for describing chemical scenarios, taking into account the effects of photons of various modes. In particular, the effect of nuclear spin photon is present, but it is much less than that of electron spin photon. If we compare Fig. \ref{fig:EffectElectronSpin} and Fig. \ref{fig:EffectNuclearSpin} obtained above, we can see that the mode $\Omega^s$ affects the association reaction much more than the mode $\Omega^n$.

	Our model is temporarily rough, but its advantage is in simplicity and scalability. And in future this model can be generalized to many modifications for laying the foundation for future complex chemical and biological models.
	
	\begin{acknowledgments}
	The reported study was funded by China Scholarship Council, project number 202108090483. The authors acknowledge Center for Collective Usage of Ultra HPC Resources (https://www.parallel.ru/) at Lomonosov Moscow State University for providing supercomputer resources that have contributed to the research results reported within this paper.
	\end{acknowledgments}

	\bibliography{bibliography}

\end{document}